\documentstyle[prd,epsfig,aps,floats]{revtex}

\newcommand{\Sslash}[1]{ \parbox[b]{0.6em}{$#1$} \hspace{-0.55em}
                         \parbox[b]{0.55em}{ \raisebox{-0.2ex}{$/$}}}
\newcommand{\beq}{\begin{equation}}
\newcommand{\eeq}{\end{equation}}
\newcommand{\beqa}{\begin{eqnarray}}
\newcommand{\eeqa}{\end{eqnarray}}

\newcommand{\lsim}{\lesssim}

\newcommand{\half}{\frac{1}{2}}

\newcommand{\ie}{{\it i.e.}}

\newcommand{\cf}{{\it cf.\ }}
\newcommand{\etal}{{\it et al.}}
\newcommand{\gev}{{\rm GeV}}
\newcommand{\mev}{{\rm MeV}}

\newcommand{\ieps}{i\epsilon\,}

\newcommand{\order}[1]{${\cal O}(#1)$}

\newcommand{\eq}[1]{Eq.\ (\ref{#1})}

\newcommand{\pvec}{\bbox{p}}

\newcommand{\kvec}{\bbox{k}}
\newcommand{\qvec}{\bbox{q}}

\newcommand{\as}{\alpha_s}
\newcommand{\lqcd}{\Lambda_{QCD}}
\newcommand{\ket}[1]{\vert{#1}\rangle}
\newcommand{\bra}[1]{\langle{#1}\vert}
\newcommand{\kmod}{\langle{k}\ket{k}}

\newcommand{\PL}[3]{Phys.\ Lett.\ {{\bf#1}} ({#3}) {#2}}
\newcommand{\NP}[3]{Nucl.\ Phys.\ {{\bf#1}} ({#3}) {#2}}
\newcommand{\PR}[3]{Phys.\ Rev.\  {{\bf#1}} ({#3}) {#2}}
\newcommand{\PRL}[3]{Phys.\ Rev.\ Lett.\ {{\bf#1}} ({#3}) {#2}}

\begin{document}

\twocolumn[\hsize\textwidth\columnwidth\hsize\csname @twocolumnfalse\endcsname
\title{%
\hbox to\hsize{\normalsize\hfil\rm NORDITA-1999/52 HE}
\hbox to\hsize{\normalsize\hfil hep-ph/9908501}
\hbox to\hsize{\normalsize\hfil August 30, 1999}
\vskip 40pt
Soft Perturbative QCD\cite{byline1}}
\author{Paul Hoyer}
\address{Nordita\\
Blegdamsvej 17, DK--2100 Copenhagen, Denmark\\
www.nordita.dk}

\maketitle

\begin{abstract}
There are indications of a faith transition in QCD: The strong coupling may
freeze at a sufficiently low value to make PQCD relevant even in
the confining regime. The properties of PQCD at low $Q^2$ depend on
the asymptotic field configuration assumed at $t=\pm\infty$. A change in the
i$\epsilon$ prescription of the free propagators is equivalent to including
quarks and gluons in the initial and final states. Perturbative expansions
thus generated are formally as justified as standard PQCD and may turn out to
be phenomenologically relevant. I discuss examples of the effects of such
propagator modifications in QED and QCD.
\end{abstract}
\vskip2.0pc]


\section{The Faith Transition} \label{sec1}

Comparisons of physical gauge theories with data rely on
perturbation theory. In particular, the successful comparisons of
perturbative QCD (PQCD) with data on hard processes has established QCD as
the correct theory of the strong interaction.

PQCD does not correctly describe soft processes, where
quarks get dressed to massive constituent quarks that are confined to
hadrons. It is commonly assumed that this physics is related to a strongly
coupled non-perturbative sector of QCD. Evidence is, however,
building up that this assumption may be wrong. Phenomenological analyses
point to a situation where the strong coupling $\as(Q^2)$ `freezes'
for low $Q^2$ at a moderate value $\as(0)/\pi= 0.14 \ldots 0.17$. The
insight that the physics of confinement may be described perturbatively has
been dubbed the `QCD faith transition' by Dokshitzer \cite{dokshitzer}.

The following examples indicate a close empirical connection between hard and
soft processes. PQCD continues to work even at low $Q^2$ provided a
regularization of the form $Q^2 \to Q^2+Q_0^2$ is made, where $Q_0$ is a
hadronic scale.
\begin{itemize}
\item {\em DIS scaling.} D. Haidt \cite{haidt} has shown that
HERA data on the structure function $F_2(x,Q^2)$ in the range $x \leq
.001$ can be fit by the simple function $F_2 = 0.41 \log(0.04/x)
\log(1+Q^2/Q_0^2)$, where $Q_0^2=0.5\ \gev^2$, see Fig. 1. This
parametrization works for all available $Q^2$, $0.1 \lsim Q^2/\gev^2 \lsim
35 $, with no sign of a PQCD to NPQCD `phase transition'.

\begin{figure}[htb]
\center\leavevmode
\epsfxsize=8cm
\epsfbox{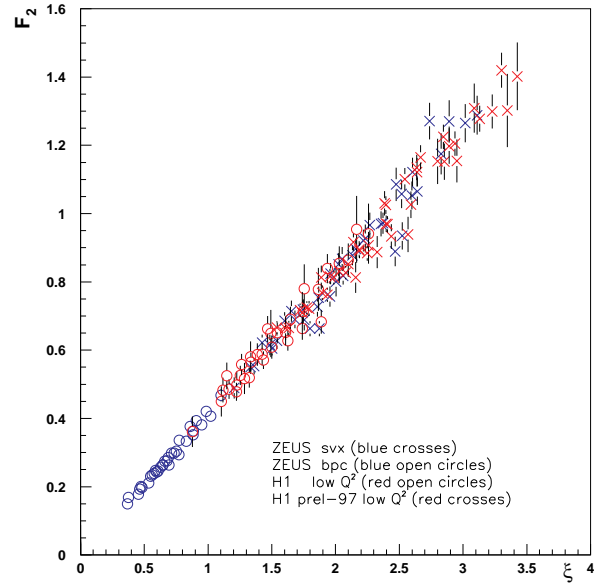}
\medskip
\caption{$F_2$ data from HERA plotted in terms of the
scaling variable $\xi=\log(0.04/x) \log(1+Q^2/Q_0^2)$. Figure
from Ref. [2].}
\end{figure}

\item {\em Exclusive scaling.} Cross sections of exclusive
processes obey dimensional scaling down to
remarkably low momentum transfers. A striking example is $\sigma(\gamma + d
\to p+n)\propto E_{CM}^{22}$ at $\theta_{CM}=89^{\circ}$ for
$E_\gamma = 1 \ldots 4$ GeV \cite{bochna} (see Fig. 2). The PQCD prediction
is proportional to $\as^{10}$ and agrees with dimensional scaling only
provided the coupling is frozen at low momentum transfers.

\begin{figure}[htb]
\center\leavevmode
\epsfxsize=8cm
\epsfbox{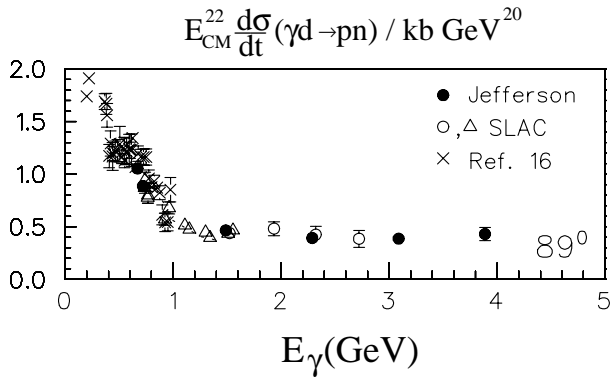}
\medskip
\caption{The $\gamma d \to pn$ cross section at $89^\circ$ multiplied by
$E_{CM}^{22}$ as a function of the photon beam energy [3].}
\end{figure}

\item {\em Local Parton Hadron Duality.} The inclusive {\em hadron}
distribution in $e^+e^- \to h+X$ and in $eN \to h+X$ appears to faithfully
track the inclusive {\em gluon} distribution calculated using PQCD, down to
momenta of \order{200\ \mev} \cite{khoze}.

\end{itemize}

Fazed by the intriguing possibility that QCD is a weakly coupled theory also
in the confinement regime it seems imperative to search for
perturbative expansions which can describe data even at low
momentum transfers. In the examples above, a hadronic scale was added to
standard PQCD `by hand', without theoretical justification.

PQCD is uniquely defined by the QCD lagrangian and by the boundary
condition at asymptotic times, $t=\pm \infty$. Formally, the choice of
boundary state $\ket{\Omega_0}$ does not affect the exact all-orders sum,
provided that it has an overlap with the true ground state
$\ket{\Omega}$:
\beq
\langle\Omega\ket{\Omega_0} \neq 0 \ \ \ \ \Longrightarrow \ \ \ 
\lim_{t\to\infty(1-\ieps)} \ket{\Omega_0,t} \propto \ket{\Omega}
\label{overlap}
\eeq

In standard PQCD we choose $\ket{\Omega_0} = \ket{0}$, the empty
`perturbative vacuum'. This choice works well in QED.
Apparently the true QED vacuum, even though extremely complicated,
is sufficiently close to $\ket{0}$ for the deviations to be treated
perturbatively. The long-distance regime of QCD is, on the other hand,
believed to be influenced by a non-trivial `gluon condensate' vacuum
\cite{condensate}. We should then consider other choices for
$\ket{\Omega_0}$, which are sufficiently simple to allow analytic
calculations yet are better models of the true QCD ground state than
the perturbative vacuum.

Standard PQCD has the elegant feature of being exactly Lorentz covariant at
each order of $\as$. S-matrix elements that have free (quark and gluon)
asymptotic states manifest their full symmetries at each order of
perturbation theory. Conversely, S-matrices whose asymptotic states are
bound (hadrons) require infinite orders of $\as$ just to generate the bound
state poles. It is unlikely that such resummations can preserve explicit
Lorentz covariance at each order of the approximation.

Bound states wave functions are defined on equal time or light-like
surfaces. As emphasized by Dirac \cite{dirac}, Lorentz boost generators that
do not leave those surfaces invariant have non-trivial representations which
involve interactions.

Consider the Fock expansion of the true QCD vacuum (disregarding quarks),
\beq
\ket{\Omega} = \phi_0 \ket{0} + \sum_{n=2}^\infty \int \left[\prod_i^n
\frac{d^3\pvec_i}{(2\pi)^3}\right]\,
\phi_n(\pvec_i) \ket{g(\pvec_1) \cdots g(\pvec_n)}
\label{omega}
\eeq
The gluon condensate contributes primarily to the long distance Fock
amplitudes $\phi_{n}(|\pvec_i| \lsim \lqcd)$. The boost invariance of the
ground state,
\beq
\hat M^{0k} \ket{\Omega} \propto \ket{\Omega}
\label{omegaboost}
\eeq
implies that the condensate appears in the low momentum Fock components
$|\pvec_i| \lsim \lqcd$ {\em in all frames}. This is possible since the
exact boost $\hat M^{0k}$ contains interactions. A free boost $\hat M_0^{0k}$
would simply translate the momenta in each Fock component of \eq{omega},
resulting in a very different state.

Describing the boost invariance (\ref{omegaboost}) of the true QCD vacuum is
a task of the same magnitude as showing that it is an eigenstate of the
exact hamiltonian. Approximate, perturbative calculations satisfy neither
exact time translation nor boost invariance. PQCD asymptotic states
$\ket{\Omega_0}$ which contain particles with momenta of \order{\lqcd} will
not be invariant even under free boosts.

The general property (\ref{overlap}) guarantees exact Lorentz covariance of
the full answer, even for non-covariant asymptotic states
$\ket{\Omega_0}$. Only the {\em rate of convergence} of the perturbative
series may be frame dependent. This can be illustrated for $\as=0$ by
choosing $\ket{\Omega_0}$ to be a superposition of the perturbative vacuum
and a state with a gluon of momentum $\kvec$,
\beq
\ket{\Omega_0} = \phi_0 \ket{0} + \phi_1 \ket{g(\kvec)}
\label{supstate}
\eeq
This state evolves in time $t$ into
\beq
e^{-iH_0t}\ket{\Omega_0} = \phi_0 \ket{0} + \phi_1
e^{-i|\kvec|t}\ket{g(\kvec)}
\label{freestatet}
\eeq
The gluon state $\ket{g(\kvec)}$ is exponentially suppressed wrt.
the perturbative vacuum $\ket{0}$ in the asymptotic time limit $t\to
\infty(1-\ieps)$.

The above arguments are formal, but so is the justification of standard
PQCD. The central point is that it is theoretically justified to consider
perturbative expansions with alternative boundary states $\ket{\Omega_0}$. A
choice different from $\ket{0}$ is motivated by the known properties of the
QCD ground state. The study and interpretation of PQCD for non-trivial
$\ket{\Omega_0}$ is theoretically fascinating and could turn out to be
phenomenologically relevant.

\section{The Perturbative Gluon Condensate} \label{pgcsec}

In considering boundary states $\ket{\Omega_0}$ that might model the gluon
condensate, it is instructive to start with perturbation theory
in a fermion condensate. Denote the free fermion propagator evaluated between
states $\ket{\bar f(\kvec,\half)\bar f(\kvec,-\half)}$ of two antifermions
with 3-momenta $\kvec$ and spin $\pm\half$ by
\beq
iS_k(x-y) \equiv
\bra{\bar f\bar f}T\{\psi(x)\bar \psi(y)\}\ket{\bar f\bar f}/\kmod^2
\label{twofxy} 
\eeq
where $\langle{k}\ket{k}=(2\pi)^3 2E_k \delta^3(\bbox{0})$. In momentum space
\beq 
S_k(p)= \left\{
\begin{array}{ll}S_F(p)&\ \ \ \ (\pvec \neq -\kvec) \\
                 S_E(p)&\ \ \ \ (\pvec = -\kvec) \\
\end{array} \right.  
\label{twofp}
\eeq 
where
\beq S_E(p)= \frac{\Sslash{p}+m}{(p^0-E_k+\ieps)(p^0+E_k+\ieps)}  
\label{sep}
\eeq 
differs from the standard Feynman propagator $S_F(p)$ only in the $\ieps$
prescription at\footnote[2]{Adding fermions rather than antifermions to the
asymptotic states would change the $\ieps$ prescription at the $p^0=+E_k$
pole.} $p^0=-E_k$. \eq{twofp} may also be expressed as
\beq
S_k(p) = S_F(p)+i\frac{\Sslash{p}+m}{\langle{k}\ket{k}} (2\pi)^4
\delta^4(p+k)
\label{fproprel}
\eeq
where $k^0=E_k$. Filling all antifermion states with $|\kvec|<\Lambda$ gives
a propagator that equals $S_E(p)$ for all $|\pvec|<\Lambda$.

The equivalence between adding fermions to the boundary states and a change
of $\ieps$ prescription generalizes to Green functions at arbitrary order
in perturbation theory \cite{pgc}:

{\em Any Green function (to arbitrary order in the coupling $\as$)
evaluated with the $S_k$ fermion propagator gives the same result as a
standard calculation using $S_F$ with antifermions of momentum $\kvec$ added
to the initial and final states.}

This statement is illustrated in Fig. 3 for the case of a single fermion
loop. Modifying the pole prescription of the free
propagator for a range of 3-momenta $\kvec$ thus allows to include
the effect of large number of `background' fermions. The apparent
non-causality of the propagation stems from an interference between the
background fermions and those propagating (on-shell) in the Green function.
The gauge invariance of the standard ($S_F$) calculation with
external fermions guarantees the gauge invariance of the calculation using
$S_k$.

\begin{figure}[htb]
\center\leavevmode
\epsfxsize=8cm
\epsfbox{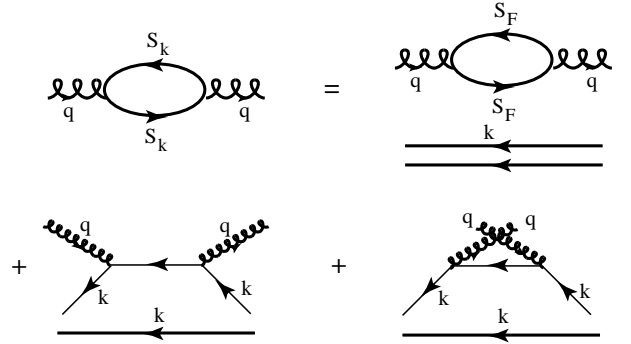}
\medskip
\caption{A fermion loop calculated with the modified propagator $S_k(p)$ 
is equivalent to a standard calculation using the Feynman
propagator $S_F(p)$ plus interactions with external antifermions of
momentum $k$.}
\end{figure}

Due to the Pauli principle a fermion condensate prevents the creation of
(anti-)fermions with the same momenta as the external ones. In a calculation
using $S_k$ this is ensured by the $\ieps$ sign change, which puts both
poles of the propagator on the same side of the real axis, thus preventing a
pinch.

It seems natural to ask whether a corresponding change of $\ieps$ sign in
{\em boson} propagators is also equivalent to standard (Feynman) perturbation
theory in the presence of external bosons. This can then serve
as a prescription for a `perturbative gluon condensate' $\ket{\Omega_0}$,
which incorporates effects of multiple gluons already at lowest
order of PQCD. In particular, such a procedure implies an exclusion
principle for bosons, preventing the emission of soft gluons.

As is often the case, the situation for bosons is analogous to that for
fermions, with a twist. For a scalar boson that is identical to
its antiparticle the Green functions do not depend on whether the sign of
$\ieps$ is changed at the positive or negative energy pole of the
propagator. Modifying the prescription at the negative energy pole for a
single momentum $\kvec$ gives a scalar propagator analogous to
\eq{fproprel}
\beq
D_k(p) = D_F(p)+\frac{i}{\langle{k}\ket{k}} (2\pi)^4 \delta^4(p + k)
\label{bproprel}
\eeq
In a realistic application one would change the prescription for a range of
momenta, say for $|\kvec|<\Lambda$.

Perturbation theory based on $D_k$ is again equivalent to the Feynman ($D_F$)
one, in the following sense:

{\em  Any Green function evaluated (at arbitrary order in the perturbative
expansion) using the modified scalar propagator $D_k$ is equivalent to a
specific superposition of Feynman Green functions with $0,1,2,\ldots$ scalars
of momentum $\kvec$ in the initial and final states.}

A derivation of this, together with the weights for each term in the
superposition, may be found in Ref. \cite{pgc}. There are only
diagonal terms, \ie, as many scalars of momentum
$\kvec$ in the incoming as in the outgoing state. The absence of cross
terms means that there is no unique boundary state $\ket{\Omega_0}$.
However, the arguments based on \eq{overlap} may be applied to each term in
the superposition separately, to show that it generates a formally correct
perturbative series. Many of the terms
will have non-vanishing total 3-momentum and hence no overlap with the true
vacuum -- their contributions will be exponentially suppressed as in
\eq{freestatet}.

Similar reasoning can be applied to QCD. The {\em perturbative gluon
condensate} (PGC) proposal \cite{pgc} is to modify the Feynman gluon
propagator at low 3-momenta, $|\pvec|<\lqcd$:
\beqa
D_\Lambda^{\mu\nu}(p) = D_F^{\mu\nu}(p)&-&\frac{i\pi g^{\mu\nu}}{2|\pvec|}
\left[\delta(p^0-|\pvec|)+\delta(p^0|+\pvec|)\right] \nonumber \\
&\times&\theta(\lqcd-|\pvec|)
\label{pgcgluon}
\eeqa
This is written in Feynman gauge with the tacit assumption that
the contributions from unphysical gluon polarizations and ghosts, which
appear as external particles in the equivalent Feynman calculation, cancel
among themselves (as they do in internal lines of standard PQCD). A
corresponding modification is then required for the ghost propagator,
\beqa
D_\Lambda^{ghost}(p) = D_F^{ghost}(p) &+& \frac{i\pi}{2|\pvec|}\,
\left[\delta(p^0-|\pvec|)+\delta(p^0|+\pvec|)\right] \nonumber \\
&\times&\theta(\lqcd-|\pvec|)
\label{pgcghost}
\eeqa
Averaging over the $\ieps$ modification at the positive and negative
energy poles ensures ghost-antighost symmetry. The same average should be
taken in the gluon propagator to satisfy Ward identities.

In the examples of the next section the propagator modifications
do, as expected, preserve gauge invariance. It would be important to have an
all orders proof of this\footnote[3]{An alternative procedure would be to
change the $\ieps$ prescription only for physical, transverse gluons (and not
for ghosts). Gauge invariance should then be guaranteed since only physical
gluons appear as external particles in the equivalent formulation in terms
of Feynman propagators.}.

\section{Examples}

The PGC prescription (\ref{pgcgluon}),
(\ref{pgcghost}) modifies PQCD at low momentum transfers. Due to
scattering on the condensate, quarks and gluons get a finite propagation
length. Some effort will be required to fully understand and interpret the
structure of the modified Green functions. In this section I shall only give
some examples of the consequences of the $\ieps$ modification in QED and QCD.

\subsection{Massless QED in 1+1 Dimensions}

The Schwinger model \cite{schwinger}, {\em alias} massless QED in 1+1
dimensions, is exactly solvable. After path integrating the electron field
the generating functional
\beqa
Z[J] &=& \int {\cal D} (A){\rm\exp}\left\{i \int d^2x
\left[-{1\over 4}F_{\mu\nu}F^{\mu\nu} \right. \right. \label{genf} \\
&-& \left. \left. {e^2\over 2\pi}A_\mu(x)\left(-g^{\mu\nu}+
{\partial^\mu\partial^\nu\over \partial^2} \right) A_\nu(x) + J_\mu A^\mu
\right] \right\} \nonumber
\eeqa
describes non-interacting photons of mass $e/\sqrt{\pi}$. The mass term is
generated by the geometric sum of one-loop corrections to the free photon
propagator shown in Fig. 4a,
\beq
\Pi_2^{\mu\nu}(q) = \frac{e^2}{\pi}\left(q^2g^{\mu\nu} - \frac{q^\mu
q^\nu}{q^2} \right)  \label{pi2}
\eeq

\begin{figure}[htb]
\center\leavevmode
\epsfxsize=8cm
\epsfbox{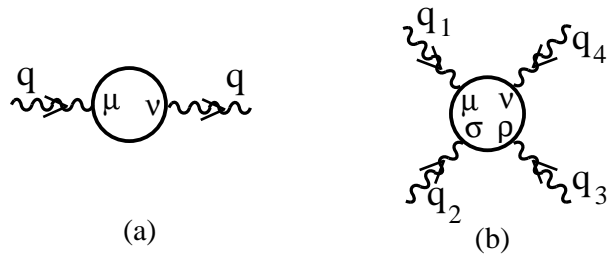}
\medskip
\caption{Electron loops in $1+1$ dimensional massless QED.}
\end{figure}

Loops with more than two external photons must vanish for the massive photons
to be non-interacting. An explicit calculation of the 4-point loop
$\Pi_4^{\mu\nu\rho\sigma}$ in Fig. 4b using LC coordinates
$q^\pm = (q^0 \pm q^1)$ shows that it is nonvanishing only for equal
$\mu=\nu=\rho=\sigma = \pm$, with
\beq
\Pi_4^{++++}= \frac{4e^4}{\pi}\left(\frac{q_1^+}{q_2^- q_{23}^- q_{234}^-}
+ \frac{q_{12}^+}{q_3^- q_{34}^- q_{341}^-} +
\frac{q_{123}^+}{q_4^- q_{41}^- q_{412}^-} \right) 
\label{pi4}
\eeq
Here $q_{ij} \equiv q_i+q_j$, {\em etc.} Summing over the permutations of the
external lines one finds that the total contribution of the fermion loop
with 4 external photons of arbitrary off-shell momenta $q_i$ ($\sum_i
q_i = 0$) indeed vanishes.

We may now ask how this situation is changed when the electron
propagator is modified as in \eq{fproprel}. For the 2-point function of Fig.
4a the change is
\beq
\delta_k\Pi_2^{++}(q) = \frac{4e^2}{\kmod}k^+\left(\frac{1}{q^- -k^-} -
\frac{1}{q^- +k^-} \right) = 0 
\label{pi2k}
\eeq
which vanishes due to $k^+k^-=0$.
The modification to the 4-point loop of Fig. 4b similarly vanishes trivially
if $k^+=0$. For $k^-=0$ it is
\beqa
\delta_k\Pi_4^{++++}&=& \frac{16e^4}{\kmod}k^+ \left(
\frac{1}{q_1^-q_{12}^-q_{123}^-} + 
\frac{1}{q_2^-q_{23}^-q_{234}^-} \right. \nonumber\\ 
&+&\left. \frac{1}{q_3^-q_{34}^-q_{341}^-} +
\frac{1}{q_4^-q_{41}^-q_{412}^-} \right)  = 0
\label{mod4}
\eeqa
The vanishing of the expression in parentheses is most easily seen from its
proportionality to the integral
\beq
\int_{-\infty}^\infty \, \frac{dx}{(x+\ieps)(x+q_1^-+\ieps)
(x+q_{12}^-+\ieps) (x+q_{123}^-+\ieps)}
\eeq
When the integration contour is closed in the lower half plane we get the
expression in \eq{mod4}. The integral vanishes since the integrand is
analytic in the upper half plane.

I conclude that (at least to \order{e^4}) the exact Schwinger model
solution is unchanged when the electron propagator is modified according to
\eq{fproprel}. This example illustrates that the non-covariant
modification (\ref{fproprel}) is compatible with fully Lorentz covariant
Green functions.

\subsection{Bound States of Weakly Coupled QED$_2$}

Two-dimensional QED simplifies also in the weak coupling limit
$m >> e$, where $m$ is the electron mass \cite{coleman}. Positronium
then has no admixture of higher Fock states, since electron
loop corrections are suppressed by $e^2/m^2$ and there are no transverse
photons. The bound state dynamics at rest is given
simply by the Schr\"odinger equation. It is, however, interesting to
consider the wave function of positronium moving at a relativistic CM
velocity.

Since the bound fermions are never on-shell one may, up to loop corrections,
alter the $\ieps$ prescription such that fermions propagate only forward
in time. This corresponds to the modification (\ref{fproprel}) for all
momenta $\kvec$. One obtains \cite{relbse} in this way a bound state equation
for the equal time wave function 
\beq
\phi_p(t;x_1,x_2) = e^{-iE_pt}\, e^{ip(x_1+x_2)/2}\, \chi_p(x_1-x_2)
\label{phieq}
\eeq
of two fermions of masses $m_{1,2}$ and CM momentum $p$ of the form
\beqa
&-&i\left[\gamma^0\gamma^1,\,\frac{\partial}{\partial x}\chi_p(x) \right] +
\frac{p}{2} \left\{\gamma^0\gamma^1,\,\chi_p(x) \right\}
\nonumber\\ 
&+& m_1\gamma^0 \chi_p(x) -m_2  \chi_p(x)\gamma^0 = 
\left[ E_p-V(x) \right] \chi_p(x)
\label{chieq}
\eeqa
where $V(x)=e^2 |x|/2$ is the Coulomb potential.

Eqs. (\ref{phieq}) and (\ref{chieq}) have no explicit Lorentz covariance,
since the wave function is defined at equal time in all frames. Thus the
dependence of the energy eigenvalue $E_p$ on the CM momentum parameter $p$ is
in general not simple. However, it turns out that precisely
(and only) for the case of a linear potential \eq{chieq} implies
$E_p=\sqrt{p^2+M^2}$, with $M$ being the positronium rest mass.

This example serves as another reminder that non-covariant
equations can have Lorentz covariant solutions. Although the
covariance of \eq{chieq} holds for arbitrary coupling, it is only in the weak
coupling limit $e/m_{1,2} \to 0$ that loop corrections are suppressed and
Eqs. (\ref{phieq}), (\ref{chieq}) can describe positronium in $1+1$
dimensions. In this limit the rest frame $(p=0)$ equation reduces to the
non-relativistic Schr\"odinger equation, and the wave function
$\chi_p(x)$ Lorentz contracts with increasing $p$.

\subsection{Scalar QED in 3+1 Dimensions}

Scalar QED offers an opportunity for studying the effect of the boson
propagator modification (\ref{bproprel}) on perturbation theory in a setting
that has many similarities with PQCD. Defining the photon self-energy
correction $\Pi(q)$ through
\beq
\Pi^{\mu\nu}(q) = \left(q^2 g^{\mu\nu}-q^\mu q^\nu \right) \Pi(q)
\label{pidef}
\eeq
the one-loop result using Feynman propagators is
\beq
\Pi_F(q) = \frac{\alpha}{8\pi}\int_0^1 dx \sqrt{1-x}
\log\left(1-\frac{q^2x}{4m^2}\right)
\label{piF}
\eeq

The modified propagator (\ref{bproprel}) gives a two-point
function $\Pi_k^{\mu\nu}(q)$ which differs from the one obtained using
Feynman propagators by
\beq
\Delta_k\Pi^{\mu\nu}(q) = 
\frac{e^2}{\kmod} \sum_{\sigma=\pm 1}\left[ \frac{(2k+\sigma q)^\mu
(2k+\sigma q)^\nu}{q^2+2\sigma k\cdot q} -g^{\mu\nu} \right]
\label{dpik1}
\eeq
The contribution from the product of $\delta$-functions has been left out
in \eq{dpik1}. At the exceptional values of $q$ where both propagators are
modified at the same loop momentum the loop integral vanishes, since all
$\ieps$'s have the same sign.

The modification (\ref{dpik1}) is readily seen to satisfy the gauge
invariance condition $q_\mu \Delta_k\Pi^{\mu\nu}(q) =0$. Without loss of
generality we may take $\qvec$ along the 3-axis,
$q=(q^+,q^-,\bbox{0}_\perp)$, with $q^{\pm}=q^0\pm q^3$. Gauge invariance
then constrains $\Pi^{\mu\nu}$ to the form (\ref{pidef}) for
$\mu,\nu = \pm$, even without assuming Lorentz invariance. One finds,
\beq
\Delta_k\Pi_L = \frac{2e^2}{\kmod} \frac{4(m^2+k_\perp^2)-q^2}{q^4-
4(q\cdot k)^2}
\label{piL}
\eeq 
where the subscript $L$ indicates that $\Delta_k\Pi_L$ refers to
Eqs. (\ref{pidef}), (\ref{dpik1}) for $\mu,\nu = \pm$ only. The transverse
($i,j=1,2$) components of \eq{dpik1} give
\beq
\Delta_k\Pi^{ij} = \frac{2e^2}{\kmod} \left(\delta^{ij}
+\frac{4k^ik^jq^2}{q^4-4(q\cdot k)^2} \right)
\eeq

There are also cross terms $\Delta_k\Pi^{\pm,i} \propto k^i$. When
summed over all directions of $\kvec$ as in the PGC
case (\ref{pgcgluon}) such terms will cancel. For the present
discussion it is sufficient to assume that an integral over the transverse
direction (azimuthal angle) of $\kvec$ is done. This implies
$\Delta_k\Pi^{\pm,i} =0$ and
$\Delta_k\Pi^{ij}= -q^2\delta^{ij} \Delta_k\Pi_T$, where
\beq
\Delta_k\Pi_T = -\frac{\alpha}{\kmod} \left[ \frac{1}{q^2}+
\frac{2k_\perp^2}{q^4-4(q\cdot k)^2} \right]
\label{piT}
\eeq

The geometric sum of the 1PI contribution (\ref{pidef}) gives the photon
propagator
\beq
\Sigma^{\mu\nu}(q) = -\frac{g^{\mu\nu}-q^\mu q^\nu/q^2}{q^2}
\frac{1}{1-\Pi(q)}
\label{sumpi}
\eeq
The usual argument that gauge and Lorentz invariance implies a vanishing of
the photon mass in perturbation theory follows from \eq{sumpi}
given that $\Pi(q)$ is finite at $q^2=0$. This is verified for Feynman
propagators from the expression (\ref{piF}).

With the modification (\ref{bproprel}) the geometric sum must be done
separately for the longitudinal and transverse components (which do not mix
as explained above). The function
$\Pi(q)$ in \eq{sumpi} then depends on the photon polarization,
\beqa
\Pi_L(q) &=& \Pi_F(q) + \Delta_k \Pi_L(q) \nonumber\\
\Pi_T(q) &=& \Pi_F(q) + \Delta_k \Pi_T(q) \label{pilt}
\eeqa
From \eq{piL} it may be seen that $\Pi_L$ is regular at $q^2=0$,
hence longitudinally polarized photons are massless. However,
$\Delta_k\Pi_T$ of \eq{piT} has a pole at $q^2=0$, implying a finite mass
for transverse photons. This is an `effective' mass, which
depends on the photon momentum $q$. Moreover, it has an imaginary part, as
we shall next discuss.

The singularities of loop integrals arise from pinches between
propagator poles that have opposite $\ieps$ prescriptions. For
Feynman propagators pinches can occur between positive and negative energy
poles, and correspond to particle-antiparticle pair production. A
change of $\ieps$ prescription at the negative energy pole as in
\eq{bproprel} removes some particle-antiparticle production -- for fermions
this ensures the Pauli exclusion principle. Instead,
`pseudothreshold' singularities arise from pinches between two negative
energy poles. These correspond to scattering of the propagating photon on
the background scalars, analogous to the lower two diagrams of Fig. 3. The
pseudothreshold singularities occur at $q^2= \pm 2q\cdot k$, \ie, at
\beq
q^+q^- = \pm (q^+k^- + q^-k^+)  \label{pseudothr}
\eeq
This condition has solutions for both $q^2>0$ and $q^2<0$, with
$q^+$ and/or $q^-$ of order $k^\pm$. {\it E.g.,} for $k^-=0$ we have $q^+ =
\pm k^+$ and no condition on $q^-$. 

In the PGC prescription (\ref{pgcgluon}) the background gluon momenta are
summed over $|\kvec|<\lqcd$ and the pinch singularities (\ref{pseudothr})
turn into cuts. The imaginary part that thus arises in the gluon propagator
reflects a finite propagation length of the gluon due to its scattering on
the background particles. The physical QCD condensate
will similarly disperse gluons. It is reasonable to assume that the true
eigenmodes of propagation in the QCD vacuum will be color singlets,
which do not suffer arbitrarily soft interactions. This is in fact what is
meant by the term `color confinement'. The perturbative gluon condensate
offers an opportunity for studying these issues.

\subsection{A QCD Ward Identity}

The effects of modifying the $\ieps$ prescription of PQCD propagators are
similar to those discussed above for scalar QED. In particular, gauge
invariance should be exact at each order in $\as$. As discussed in Section
\ref{pgcsec}, in covariant gauges it is important to verify that the
unphysical polarizations of gluons and ghosts cancel also for particles in
the perturbative condensate.

In Feynman gauge the modified gluon and ghost propagators are, following the
PGC prescription (\ref{pgcgluon}), (\ref{pgcghost}) for a {\em single}
momentum $k=(|\kvec|,\kvec)$,
\beqa
D_k^{\mu\nu,ab}(p) &=& -g^{\mu\nu}\delta^{ab} \left[ \frac{1}{p^2+\ieps} 
+\delta_k \right] \nonumber\\ 
D_k^{ghost,ab}(p) &=& \delta^{ab} \left[ \frac{1}{p^2+\ieps} +\delta_k
\right]
\label{qcdprop}
\eeqa
where
\beq
\delta_k = \frac{i}{2\kmod} (2\pi)^4 \left[ \delta^4(p-k) + \delta^4(p+k)
\right]
\label{modk}
\eeq

The gluon + ghost loop modification to the gluon propagator is
\beqa
&&\Delta_k\Pi_g^{\mu\nu,ab}(q) = 
\frac{g^2C_G\delta^{ab}}{\kmod}  \label{dpikg} \\
&\times& \sum_{\sigma=\pm 1}\left[ \frac{(2k+\sigma
q)^\mu (2k+\sigma q)^\nu +2(g^{\mu\nu}q^2-q^\mu q^\nu)}{q^2+2\sigma k\cdot q}
-g^{\mu\nu} \right] \nonumber
\eeqa
where as in \eq{dpik1} I have ignored the contribution from the product of
$\delta$-functions, which only contributes at discrete values of $q$.

QCD Ward identitities follow from the requirement that the BRS
transformation of Green functions vanish \cite{sterman}. The
two-point function $\bra{0}T\{\bar c_a(x_1) A_b^\mu(x_2)\}\ket{0}$, where $c$
and $A$ are ghost and gluon fields, respectively, yields the Ward identity
\beqa
&&\bra{0}T\left\{ \frac{\partial}{\partial x_1^\nu} A_a^\nu(x_1)
A_b^\mu(x_2) \right\}\ket{0}  \label{wardid} \\
&=& \bra{0}T\left\{ \bar c_a(x_1) \left[ \frac{\partial}{\partial x_{2\mu}}
\delta_{bc} + gf_{bcd} A_d^\mu (x_2) \right] c_c(x_2) \right\}\ket{0}
\nonumber 
\eeqa
At \order{g^0}, this is a relation between the free gluon and ghost
propagators which is trivially satisfied due to their identical
modification $\delta_k$ in (\ref{qcdprop}). At \order{g^2}, the lhs of
\eq{wardid} vanishes, $q_\mu \Delta_k\Pi_g^{\mu\nu,ab}(q) =0$
is readily verified from \eq{dpikg}. It turns out that the two terms on
the rhs of \eq{wardid} cancel. Hence the Ward identity remains valid
when evaluated using the modified propagators (\ref{qcdprop}).

While this check of QCD gauge invariance is non-trivial, it does not obviate
the need for a general proof.

\section{Summary}

My proposal to consider modifications of the $\ieps$ prescription in PQCD is
unusual. There seems, however, to be good reasons for pursuing such an
approach.
\begin{itemize}
\item Perturbation theory is a central tool for comparing physical gauge
theories with data. The properties of formally equivalent perturbative
series, obtained using different asymptotic states at $t=\pm\infty$, thus
merit investigation.
\item There are phenomenological indications that $\as(Q^2)$ freezes
at low $Q^2$. PQCD appears to be relevant even in the confining,
long-distance regime.
\item Relativistic corrections are indistinguishable from other higher order
effects for bound states, since $\as=v/c$. Exact Lorentz covariance is not
required at each order of the approximation \cite{caswell}.
\item The PGC prescription (\ref{pgcgluon}) provides a
gauge-invariant framework for studying the influence of background color
fields on quark and gluon interactions. All short distance properties of PQCD
are preserved.
\end{itemize}

The QCD ground state is believed to have also a quark condensate,
related to the breaking of chiral symmetry \cite{condensate}. It may thus be
interesting to consider asymptotic states $\ket{\Omega_0, t=\pm \infty}$
which contain also $q\bar q$ pairs, as a `Perturbative Quark Condensate'
model of the true vacuum. The free fermion propagator between states which
have a fermion-antifermion pair of momentum $\kvec$ differs from the Feynman
propagator in the $\ieps$ prescription at {\em both} the positive and
negative energy poles (\cf the discussion at the beginning of Section
\ref{pgcsec}). An analysis of the type outlined above for gluons should thus
be carried out also for quarks.

\vspace{1cm}

{\bf Acknowledgements} I am grateful to the organizers for their invitation
to this interesting workshop, and for helpful discussions with S. J. Brodsky,
G. W. Carter, C. S. Lam and H. Weigert.

\end{document}